# [1]Report on the piezoelectric frequency effect in quartz and its possibility in other piezoelectric materials

Yubin Hou[1], Qingyou Lu[1,2,*]

[1]*High Magnetic Field Laboratory, Chinese Academy of Science, Hefei, Anhui 230031, P. R. China*

[2]*Hefei National Laboratory for Physical Science at Microscale, University of Science and Technology of China, Hefei, Anhui 230026, P. R. China*

*Author to whom correspondence should be addressed. Tel:86-551-6360-0247. E-mail:qxl@ustc.edu.cn

**Abstract**

We report the eigen-frequency shift induced by an applied voltage on quartz material. The samples are in the form of commercial 32 KHz quartz tuning forks used in watches. Three vibration modes are studied: one prong oscillates, two prongs oscillate in the same and opposite directions. They all show a parabolic dependence of the eigen-frequency shift on the bias voltage applied across the fork, which is explained owing to the voltage-induced internal stress that varies as the fork oscillates. Thus, this piezoelectric frequency effect is possible to exist in all piezoelectric materials. The average coefficient of the piezoelectric frequency effect is as low as several hundred nano-Hz per millivolt, implying that the most precise (nano-Hz) yet fast-response voltage-controlled oscillators and phase-locked loops can be built.

**I. Introduction**

The piezoelectric (PE) effect (including the reverse PE effect) refers to the approximate linear response of the mechanical and the electrical mutual interaction in crystalline materials lacking inversion symmetry [1]. The importance of the PE effect is well known. One type of its famous application utilizes the dimension change of the PE material under applied voltage to produce the voltage controlled motion or positioning in real space. We can therefore call this type of PE effect the space domain PE effect. Its application examples are nanopositioners [2]-[3], piezoelectric motors [4]-[6], precise focusing of modern optical assemblies [7]-[8] and atomically and sub-atomically resolved scanning probe microscopes [9]-[16], etc. An obvious reason for these important applications is that the coefficient of the PE deformation is very low, around 0.1 microns per volt (or 1 Å per millivolt), providing very precise yet fast enough control of displacement.

---

1. We have done this work in March-2011 and submitted to Ultrasonic, Ferroelectrics and Frequency Control, IEEE Transactions on 19-Jul-2011.

Another particularly important application type of the PE effect is the precise measurement of time, which exploits the ultrahigh stability of the eigen-frequency of a PE oscillator. Examples include quartz tuning forks in watches, crystal oscillators in computers, quartz crystal microbalances [17]-[20], voltage controlled oscillators (VCO), phase lock loops (PLL) [21], and cantilever excitors of atomic force microscopes [22]-[26], etc. However, the precise control or adjustment of the eigen-frequency of a PE oscillator is always an issue. Traditionally, this is done by adjusting an external capacitor (which reduces the frequency stability and does not provide a high adjustment precision either) or adding mass to the oscillator (which is not a controllable method). We surely hope that an applied voltage on the PE material can cause its eigen-frequency to change in a manner of high precision and fast response. This frequency domain PE effect could be as important as the space domain PE effect mentioned above, but nevertheless it has not been reported before.

Presented in this paper is the first report on the voltage induced frequency domain PE effect, which will hereafter be referred to as piezoelectric frequency (PEF) effect for simplicity. Its coefficient is studied also and found to be easily smaller than 1 mHz per volt (or sub micro-Hertz per millivolt). This extremely low value implies its immediate important applications in the high precision versions of VCO and PLL. Other considerable applications include: precise frequency sweeper, adjustable timing devices and microbalances, etc. Also presented in this paper is a fairly good theoretical explanation of the PEF effect, which turns out not owing to the space domain PE effect induced dimension change (i.e. capacitance change) of the quartz oscillator, but due to the voltage induced internal stress that varies as the sample oscillates. This also tells that the PEF effect should exist in other (maybe all) PE materials.

## II. Experimental setup

Figure 1 shows the schematic diagram of the setup employed to measure the PEF effect. The quartz sample is in the form of quartz tuning fork (QTF, type E158 from Nanosurf AG) as used in watches. One prong of the QTF is glued by Torr Seal on a sapphire piece of L shape which is in turn mounted (by the same glue) on a dither piezo plate. This type of the QTF with one prong being fixed is often called qPlus QTF [27]. Another sapphire piece is sandwiched between the dither piezo and the base for good isolation. The oscillation details of the free prong of the QTF are measured by a preamp which is a current to voltage converter (consisting of a 100 MOhm feedback resistor wired across a Texas Instruments OPA627 operational amplifier) with its input being connected to the corresponding lead of the QTF [28]. When the free prong oscillates, the charge on it (due to the space domain PE effect) will change. The resultant current is converted into a voltage signal $V_{prong}$ by the preamp and then connected back to the driving dither piezo via an automatic gain control (AGC, which is inside the commercial Easy PLL plus controller by Nanosurf, Swiss) to form a close loop oscillation circuit. The purpose of the AGC is to maintain a constant oscillation amplitude.

A bias voltage (-130 to +130 V adjustable, generated from an RHK SPM100 scanning probe microscope controller) can be applied to the QTF by applying it across the virtual ground of the preamp and the remaining lead of the QTF. The oscillation loop ensures that the oscillation frequency tracks the QTF's instant eigen-frequency, therefore we can measure the QTF's dynamic eigen-frequency (and how it changes) by measuring the oscillation frequency of the circuit loop with a PLL (Easy PLL plus of Nanosurf, Swiss). It is important to note that the change of the eigen-frequency of the QTF can not be obtained by continuously measuring its resonance curves because each resonance curve takes long time to measure (several minutes), during which thermal drifts will reduce the accuracy of the measurement significantly.

In the setup described above, only one prong of the QTF can oscillate. We can also vibrate both prongs in the same direction by using the dither piezo to vibrate the portion of the QTF where the two prongs join (i.e. by mounting the joint on the dither piezo, not any of the prongs). This type of the mounted QTF is hereafter called in-phase QTF for simplicity. There is also the third vibration mode in which both prongs vibrate in the opposite direction (hereafter called anti-phase QTF). To this end, the setup is similar to that of the in-phase QTF except that the dither piezo is removed and its driving signal is changed to connect to the non-inverting input of the preamp (the joint of the prongs is fixed on the base).

In this paper, the qPlus, in-phase and anti-phase vibration modes are all studied and compared.

### III. Results and theoretical explanation

The measured dependences of the eigen-frequency shifts of the qPlus, in-phase and anti-phase QTF on the bias voltage applied on the QTF are shown in Fig.2, each of which excellently fits a quadratic function. The fitted parabolic curves approximately pass through the origin, so they can be written in the form $y = ax^2$ with a = 0.0039, 0.0066, and 0.0064 mHz/V$^2$, respectively. Since the eigen-frequency shift is not linearly dependent of the bias voltage, the piezo-frequency coefficient is not a constant. Nevertheless, we can still discuss its average piezo-frequency coefficient defined as eigen-frequency change divided by bias voltage change.

The so defined average piezo-frequency coefficients are extremely low and are 470, 917 and 826 nano-Hz per millivolt for the qPlus, in-phase and anti-phase QTFs, respectively, implying that the most precise (nano-Hz precision) VCOs and PLLs can be built based on this PEF effect. The average piezo-frequency coefficient can be greatly changed further by changing the geometry of the oscillator, just like changing the size of a piezoelectric actuator can change the displacement per unit voltage it generates.

To check the repeatability of the PEF effect of the three QTF modes, we mechanically switched the bias voltages on the three types of QTF repeatedly between 0 and +130 V and watch how the output signals of the Easy PLL plus look like. Its turns out that they all look similar. As an example, the step response of the anti-phase QTF is shown in Fig. 3. The repeatability is high since the high and low outputs are almost identical. The big fluctuations between high and low outputs are apparently due to the mechanical switching.

All the experiments were done in low vacuum and room temperature.

The resonance curve for the qPlus QTF is shown in Fig. 4, which is rather symmetric. The curves are a little skewed for the in-phase and anti-phase QTFs. The Q values are all above 3000 which is typical for the QTFs in low vacuum and room temperature, showing that the QTFs are well functioned.

To explain why the PEF effect can happen, we notice that applying a bias voltage V on the QTF will produce an electrostatic force $F_V$ on the QTF just as applying a voltage on a capacitor will exert an electrostatic force on the dielectric. This force $F_V$ is given by $F_V = \frac{1}{2}CV^2/d$, where C is the capacitance of the QTF in its model and d is the effective electrode separation of the capacitor C. Before the bias voltage is applied, the free oscillation (presumably in $x$ direction) of the QTF experiences only a restoring force of $F_0 = -k_0 x$. After the bias voltage is turned on, the force on the QTF becomes $F_1 = -k_0 x + F_V = -k_0 x + \frac{1}{2}CV^2/d = -k_0 x + \beta V^2/d^2$, where C is assumed to be proportional to $d^{-1}$ and $\beta$ contains all the constants in the last term. Assume d varies in $x$ direction when the QTF oscillates, then $F_1$ can be Taylor expanded as $F_1 = -k_0 x + \beta V^2/d_0^2 - x \cdot 2\beta V^2/d_0^3 + \ldots$. Thus, the force constant dynamically becomes $k_0 + 2\beta V^2/d_0^3$, which is larger than the original oscillation force constant $k_0$, leading to a larger eigen-frequency. Furthermore, the increase in the eigen-frequency is parabolically dependent of the applied voltage, which is exactly our measured result.

From this interpretation, we also know that response of the frequency domain PE effect should be as fast as the space domain PE effect because they stem from the same cause. The above explanation does not require that the material be quartz or in the shape of fork. Therefore, the PEF effect should be rather general and exist in all piezoelectric materials and in any shapes.

## IV. ACKNOWLEDGMENTS

This work was supported by the Science Foundation of The Chinese Academy of Sciences under grant number YZ200846 and the project of Chinese national high magnetic field facilities.

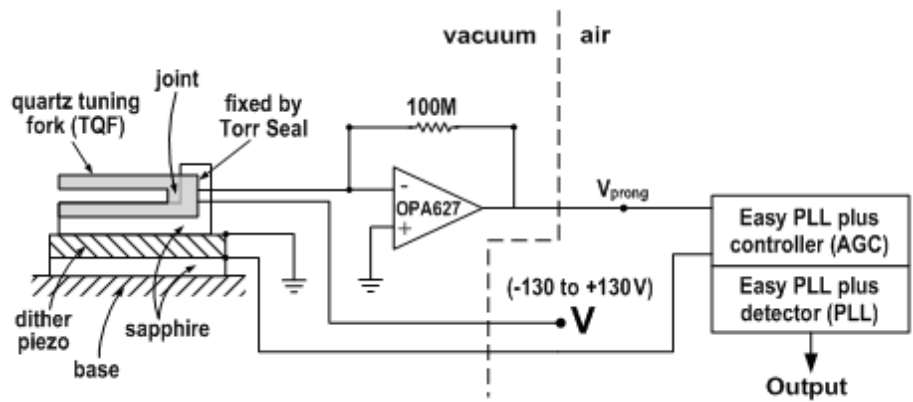

Fig.1. The schematic diagram of the setup employed to measure the PEF effect.

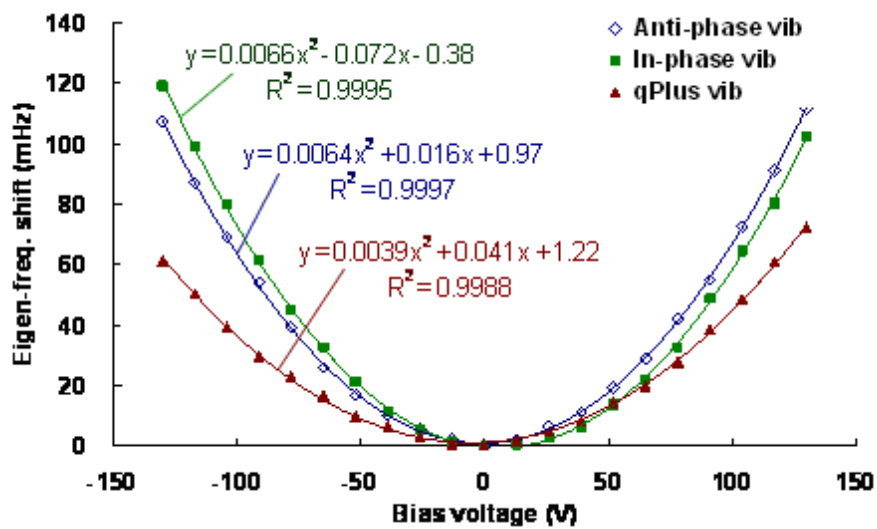

Fig.2. Plots of the eigen-frequency shifts vs. the applied bias voltages for the three different prong vibration modes. The data of each vibration mode well fit a parabolic relation. See the text for explanation.

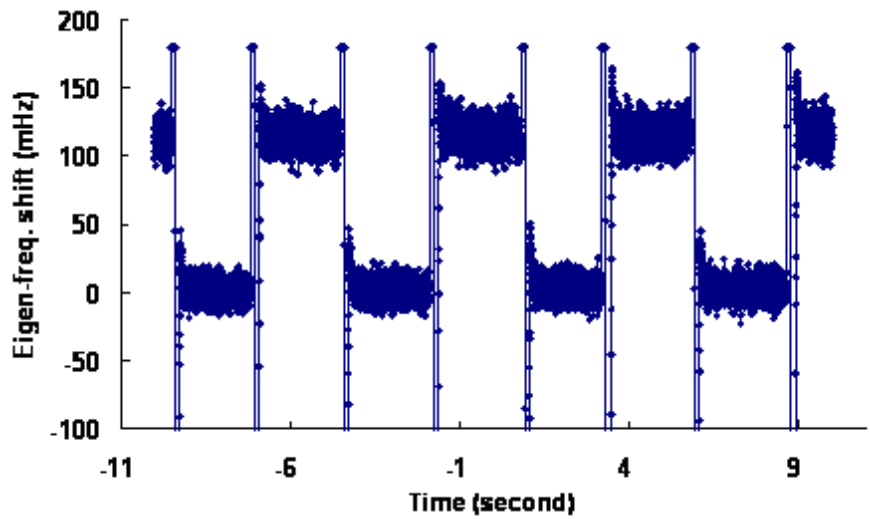

Fig.3. Plot of the eigen-frequency change vs. the repeated switch of the applied bias voltage between 0 and +130 V for the anti-phase QTF. The results are very similar for the in-phase and qPlus QTFs.

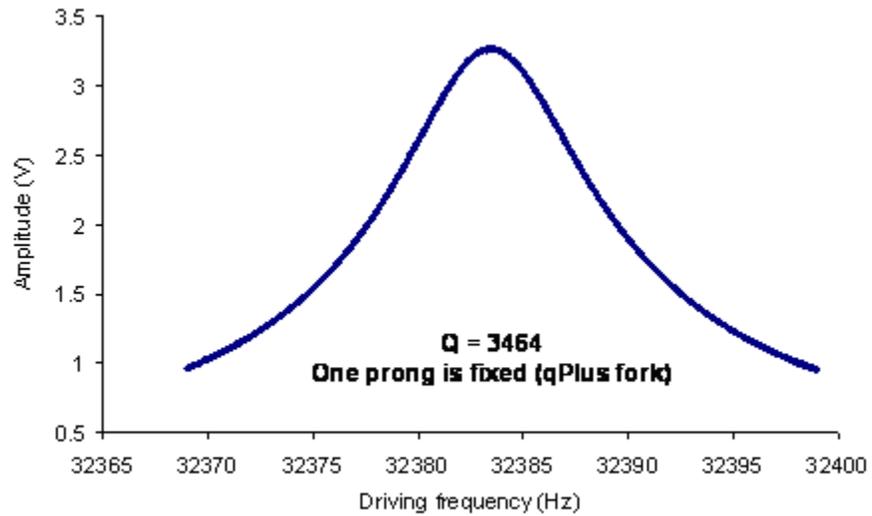

Fig.4. The resonance curve of the qPlus QTF used in the experiment. It is fairly symmetric with a Q factor equal to 3464.